\documentclass[conference,compsoc,letterpaper]{IEEEtran}
\IEEEoverridecommandlockouts
\usepackage{cite}
\usepackage{amsmath,amssymb,amsfonts}
\usepackage{algorithm, algorithmic}
\usepackage{graphicx}
\usepackage{textcomp}
\usepackage{xcolor}
\usepackage{stfloats}
\usepackage{multirow}
\usepackage{url}

\usepackage{array}
\usepackage{mdwmath}
\usepackage{mdwtab}
\usepackage{pifont}
\def\BibTeX{{\rm B\kern-.05em{\sc i\kern-.025em b}\kern-.08em
    T\kern-.1667em\lower.7ex\hbox{E}\kern-.125emX}}
\begin{document}

\title{BaPipe: Exploration of Balanced Pipeline Parallelism for DNN Training \\
\thanks{\IEEEauthorrefmark{4}Xi Jin (jinxi@ustc.edu.cn) is the corresponding author.}
}

\author{
\IEEEauthorblockN{Letian Zhao\IEEEauthorrefmark{1}\IEEEauthorrefmark{2}, Rui Xu\IEEEauthorrefmark{1}\IEEEauthorrefmark{2}, Tianqi Wang\IEEEauthorrefmark{1}\IEEEauthorrefmark{2}, Teng Tian\IEEEauthorrefmark{1}\IEEEauthorrefmark{2}, Xiaotian Wang\IEEEauthorrefmark{1}\IEEEauthorrefmark{2}, Wei Wu\IEEEauthorrefmark{1}\IEEEauthorrefmark{2}, Chio-In IEONG\IEEEauthorrefmark{3}, and Xi Jin\IEEEauthorrefmark{1}\IEEEauthorrefmark{2}\IEEEauthorrefmark{4}}
\IEEEauthorblockA{
\textit{\IEEEauthorrefmark{1}State Key Laboratory of Particle Detection and Electronics, University of Science and Technology of China, Hefei, China} \\
\textit{\IEEEauthorrefmark{2}Institute of Microelectronics, Department of Physics, University of Science and Technology of China, Hefei, China}\\
\textit{\IEEEauthorrefmark{3}Huawei Technologies, Shenzhen, China}\\
Emails: \{zhaolt, xray, tqwang, tianteng, wxtdsg, wuw1993\}@mail.ustc.edu.cn,\\
ieong.chio.in@huawei.com, jinxi@ustc.edu.cn
}
}

\maketitle

\begin{abstract}
The size of deep neural networks (DNNs) grows rapidly as the complexity of the machine learning algorithm increases. To satisfy the requirement of computation and memory of DNN training, distributed deep learning based on model parallelism has been widely recognized. We propose a new pipeline parallelism training framework, BaPipe, which can automatically explore pipeline parallelism training methods and balanced partition strategies for DNN distributed training. In BaPipe, each accelerator calculates the forward propagation and backward propagation of different parts of networks to implement the intra-batch pipeline parallelism strategy. BaPipe uses a new load balancing automatic exploration strategy that considers the parameters of DNN models and the computation, memory, and communication resources of accelerator clusters. We have trained different DNNs such as VGG-16, ResNet-50, and GNMT on GPU clusters and simulated the performance of different FPGA clusters. Compared with state-of-the-art data parallelism and pipeline parallelism frameworks, BaPipe provides up to 3.2x speedup and 4x memory reduction in various platforms.
\end{abstract}


\section{Introduction}
In recent years, deep neural networks (DNNs) have made great progress in a variety of applications such as image classification\cite{he2016deep,krizhevsky2017imagenet,he2016identity,simonyan2014very} and natural language processing (NLP)\cite{wu2016google,vaswani2017attention,devlin2019bert}. As the effect of DNNs has been improved, the size of DNNs has become bigger and bigger, making DNNs more computationally intensive and memory intensive. The adoption of a single accelerator for DNN training will face the memory bottleneck and cost unaffordable training time. Therefore, DNN training usually deploys distributed computing systems which consist of multiple accelerators such as GPUs, FPGAs, and ASICs. GPU clusters have been used most for DNN training\cite{narayanan2019pipedream,huang2019gpipe,pal2019optimizing,transientcloud2019guo,capes2019dynamic} as they are easy to implement and have high computing power. FPGA clusters have become a popular option due to the advantages of energy efficiency and flexibility\cite{geng2018fpdeep,wang2020fpdeep,venkataramani2017scaledeep,geng2018framework}. ASIC clusters are developed by many companies\cite{huang2019gpipe,fedus2021switch} for the most energy efficiency but the highest development cost at the same time. And heterogeneous clusters, such as clusters consisting of mixed models of GPUs, are commonly employed in realistic scenarios\cite{li2020characterizing,HPDL2019,zhao2019dynamic,Falcon,icdcs2019autoconfig}.

\begin{figure}[tbp]
\centerline{\includegraphics[width=\linewidth]{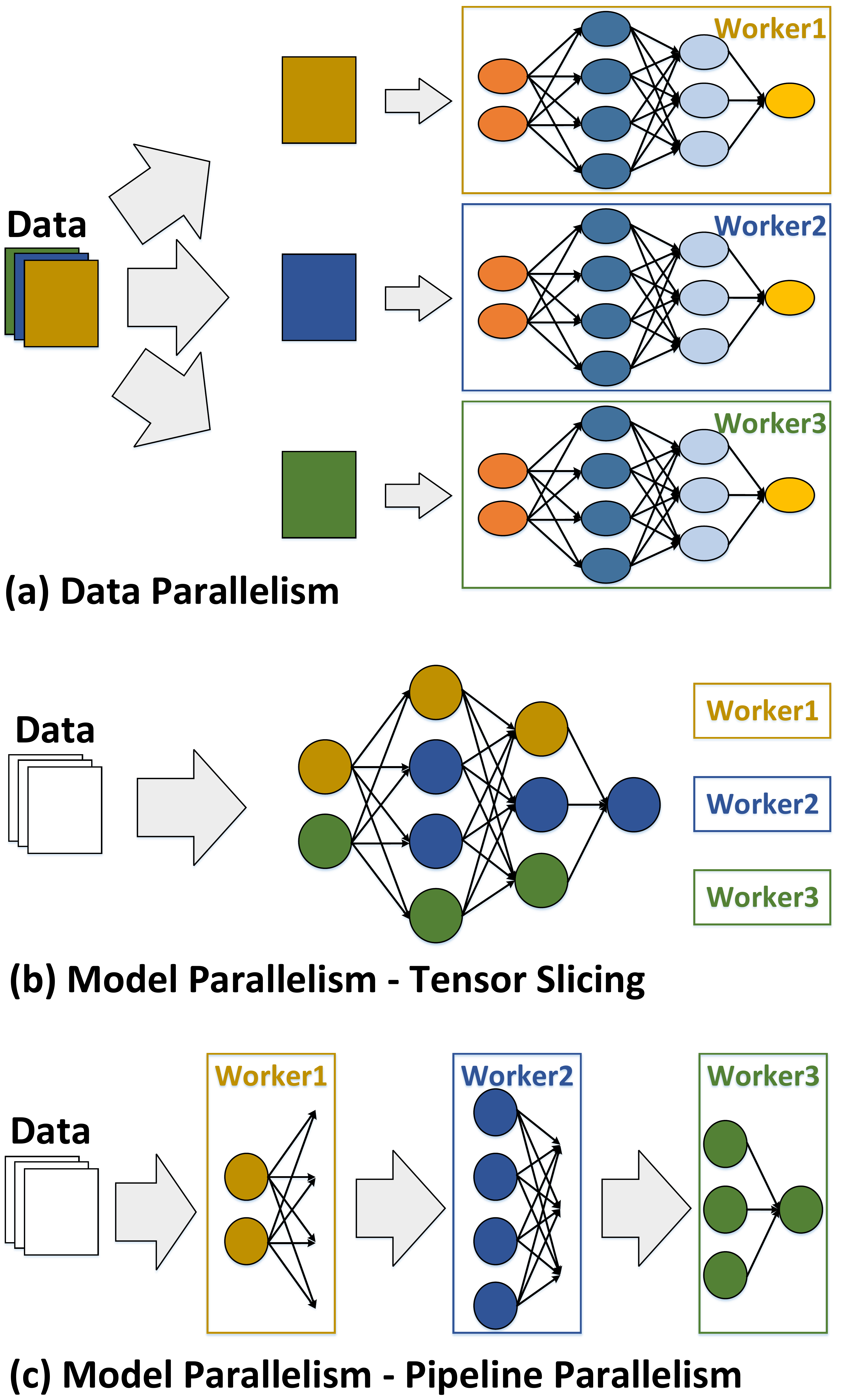}}
\caption{Different neural network training Parallelism schemes: (a) Data Parallelism, (b) Model Parallelism - Tensor Slicing, (c) Model Parallelism - Pipeline Parallelism. }
\label{parallelism}
\end{figure}

Nowadays most accelerator clusters for DNN training work in Data Parallelism\cite{ben2019demystifying,mayer2020scalable}. As shown in Fig. \ref{parallelism}(a), each accelerator in distributed systems with Data Parallelism computes the whole layers of DNN with a different part of training data concurrently. In the meantime, all the weights and intermediate features of DNN must be stored locally by each accelerator in Data Parallel systems. Therefore, the capacity of higher bandwidth memory of accelerators maybe not enough for big DNN training which will consequently decrease the parallel performance. The higher bandwidth memory is compared with low bandwidth memory in distributed systems. For example, the bandwidth of on-chip memory is higher than off-chip memory in FPGA clusters. And graphic memory is higher bandwidth memory compared with host memory in GPU clusters. Distributed data parallel systems even cannot training larger DNN models when the memory consumption of deep learning is much larger than the memory capacity of accelerators. Thus, Model Parallelism has been used in DNN training. Large DNN models have been divided into partitions. Different accelerators execute different partitions, and store different parts of weights and intermediate features as shown in Fig. \ref{parallelism}(b)(c). Hence, distributed systems with Model Parallelism can train larger models and increase training speed.

There are two types of Model Parallelism, which we called Tensor Slicing and Pipeline Parallelism. As shown in Fig. \ref{parallelism}(b), Tensor Slicing splits each layer of the DNN model horizontally across accelerators, i.e. workers in Fig. \ref{parallelism}, without pipeline. And there will be lots of All-Reduce operations among workers after every layer, which will lead to high communication overhead across workers. On the other hand, Pipeline Parallelism splits the DNN model vertically into different subsets of neural network layers, as shown in Fig. \ref{parallelism}(c). And each worker computes different partition in the pipeline, which increases parallel efficiency. There are also two approaches of Pipeline Parallelism: inter-batch and intra-batch pipeline parallelism. As shown in Fig. \ref{intra&inter}(b), the inter-batch parallelism sets one mini-batch of training data as a primitive element in the pipeline. Consequently, the weights used for backward propagation (BP) of DNN training are not the same version of weights in forward propagation (FP). Thus, it prolongs the convergent rate of a DNN model and even diverges the training of the model. In contrast, intra-batch parallelism, as shown in Fig. \ref{intra&inter}(a), divides a mini-batch into micro-batches for pipeline parallelism to achieve the high accuracy results in DNN training without inconsistent weights.

However intra-batch pipeline parallelism still faces two major challenges. First, the computation for DNN training is consistent with forward propagation (FP) and backward propagation (BP). And BP should be executed after FP with the same micro-batch of intermediate features. Thus there are various scheduling methods of the pipeline consisting of different FP and BP from different micro-batches. The different pipeline scheduling will generate varying memory requirements to store the different sizes of intermediate features, and affect parallel performance due to the different rate of pipeline bubbles. Second, the efficiency of pipeline parallelism is determined by the bottleneck of distributed training systems. That means the partition of DNN training should make the load across accelerators as balanced as possible. Communication between accelerators, including both intermediate feature maps in FP and errors in BP,  and computation costs constitute the executing time of each accelerator. In addition, the overlap of communication and computation will also have a significant impact on parallel efficiency.

To solve the challenges described above, we propose BaPipe, a novel distributed DNN training framework with intra-batch Pipeline Parallelism. BaPipe can improve DNN training performance and reduce memory consumption by automatically exploring pipeline scheduling and load balanced partition strategies. The main contributions of this work are as follows:

\begin{itemize}
\item We present an automatic exploration method of the scheduling of pipeline parallelism which can select the most suitable scheduling method according to hardware resources of accelerator clusters and structure of DNNs.
\item We propose a partitioning strategy suitable for a variety of DNNs. The DNN model is specified as consecutive layers to balance the computational load, communication cost, and memory consumption among accelerator clusters. And the partitioning strategy also applies to distributed systems in heterogeneous environments, where the clusters consist of mixed models of hardware, including GPU clusters and FPGA clusters.
\item We evaluate BaPipe by training different DNN models such as VGG-16, ResNet-50, and GNMT on various GPU clusters and simulating the performance of both homogeneous and heterogeneous FPGA clusters. Compared with state-of-the-art data parallelism and pipeline parallelism frameworks, BaPipe provides 3.2x speedup and 4x memory reduction on various platforms.
\end{itemize}

The rest of this paper is organized as follows: Section II discusses background and related work. Section III presents the framework of BaPipe. BaPipe is evaluated in Section IV and Section V concludes this paper.

\section{Background and Related Work}
Many works focus on parallel DNN training in recent years which can be classified into Data Parallelism, Model Parallelism, and Hybrid Parallelism.

\subsection{Data Parallelism}
The whole DNN model weights are replicated on each accelerator and input data of mini-batches are partitioned evenly across accelerators in Data Parallelism, as shown in Fig. \ref{parallelism}(a), thus each accelerator synchronizes updates weights after backward propagation. A popular data parallel DNN training method for large-scale distributed systems is parameter server framework\cite{icdcs2018fresh,icdcs2018shmcaffe}. To decrease communication costs between the parameter server and other accelerators, Ring All-Reduce options are used commonly nowadays\cite{goyal2017accurate,capes2019dynamic}. In this paper, we use synchronized All-Reduce Data Parallelism as a baseline.

\begin{figure*}[tbp]
\centerline{\includegraphics[width=\linewidth]{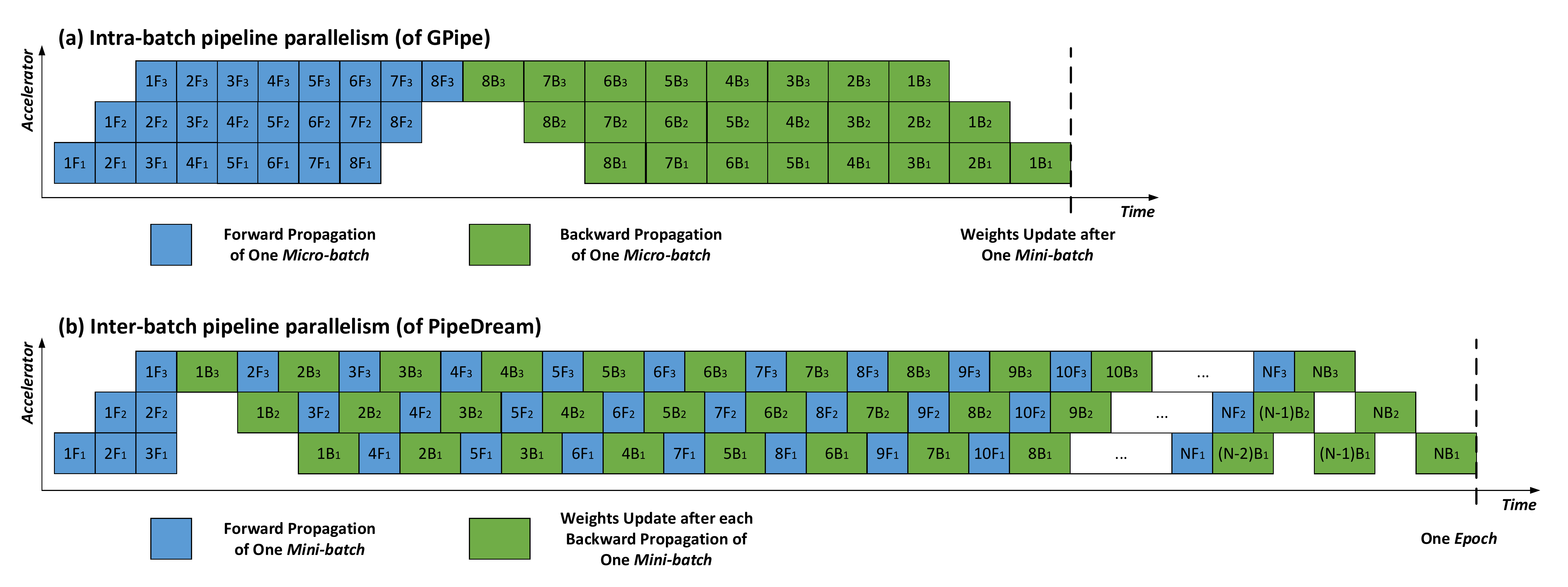}}
\caption{Different neural network training Pipeline Parallelism schemes: (a) Intra-batch Pipeline Parallelism (of GPipe), (b) Inter-batch Pipeline Parallelism (of PipeDream).}
\label{intra&inter}
\end{figure*}

\subsection{Model Parallelism}
In Model Parallelism, the DNN model is divided into partitions mapping to accelerators, thus clusters can be used in larger DNN training. There are also two ways for Model Parallelism: Tensor Slicing and Pipeline Parallelism, as shown in Fig. \ref{parallelism}(b)(c), which respectively partition models horizontally and vertically. Mesh-Tensorflow\cite{shazeer2018mesh} and Megatron-LM\cite{shoeybi2019megatron} both implemented large NLP models by Tensor Slicing. However, accelerators in Tensor Slicing have to communicate intermediate features after each layer which will degrade training efficiency a lot.

\subsubsection{Pipeline Parallelism}

Pipeline Parallelism can be divided into two paradigms according to the primitive element in the pipeline: inter-batch pipeline parallelism and intra-batch pipeline parallelism, as shown in Fig. \ref{intra&inter}(a)(b), which updates weights after each mini-batch asynchronously and synchronously, respectively. That means each accelerator in an intra-batch pipeline parallel cluster has the same fresh version of weights, at the same time, the weights of accelerators in inter-batch pipeline parallelism may be delayed or in different versions. 

PipeDream\cite{narayanan2019pipedream} is the most cited inter-batch pipeline parallelism technique in GPU clusters. The main contribution of PipeDream is weight stashing which keeps multiple versions of weights to solve the weight inconsistency problem of inter-batch pipeline parallelism. However, more memory requirements are generated by weight stashing thus larger DNN cannot train with PipeDream while the weight staleness issue still exists which impacts the convergence and accuracy of DNNs. PipeDream also presents a dynamic programming algorithm to balance computational load and communication costs across layers but ignore memory requirements. GPipe\cite{huang2019gpipe} and FPDeep\cite{geng2018fpdeep,wang2020fpdeep} are popular intra-batch pipeline parallelism approaches. GPipe split each mini-batch into micro-batches to increase pipeline training efficiency with GPUs or TPUs while backward propagation is computed after the whole forward propagation of each mini-batch being completed which leads to significant memory requirement for intermediate features of all micro-batches in one mini-batch. So GPipe sacrifices performance to reduce memory cost by recomputing forward propagation. Moreover, GPipe has not proposed a partitioning strategy to balance the load. FPDeep presents a fine-grained pipeline for FPGA clusters which partitions model intra-layer to implement high parallelism and utilization. Meanwhile, FPDeep proposes a novel workload and weight partitioning schemes but has not addressed the communication bottleneck.

\subsection{Hybrid Parallelism}
As model parallelism and data parallelism are orthogonal, there are various works used hybrid parallelism\cite{geng2019horizontal}, for example, FlexFlow\cite{jia2019beyond,jia2018exploring} automatically optimizes parallelism across the sample, operation, attribute, and parameter dimension. Megatron-LM\cite{shoeybi2019megatron} and PipeDream\cite{narayanan2019pipedream} also use hybrid parallelism. However hybrid parallelism with automatic partition has much more requirements on the topology of clusters in which arbitrary accelerators may need to be connected. Our design, BaPipe, is suitable for accelerator clusters in 1D daisy chain topology and orthogonal to data parallelism.

\begin{figure}[bp]
%\centerline{\includegraphics{fig/bapipe-framework.pdf}}
\centerline{\includegraphics[width=\linewidth]{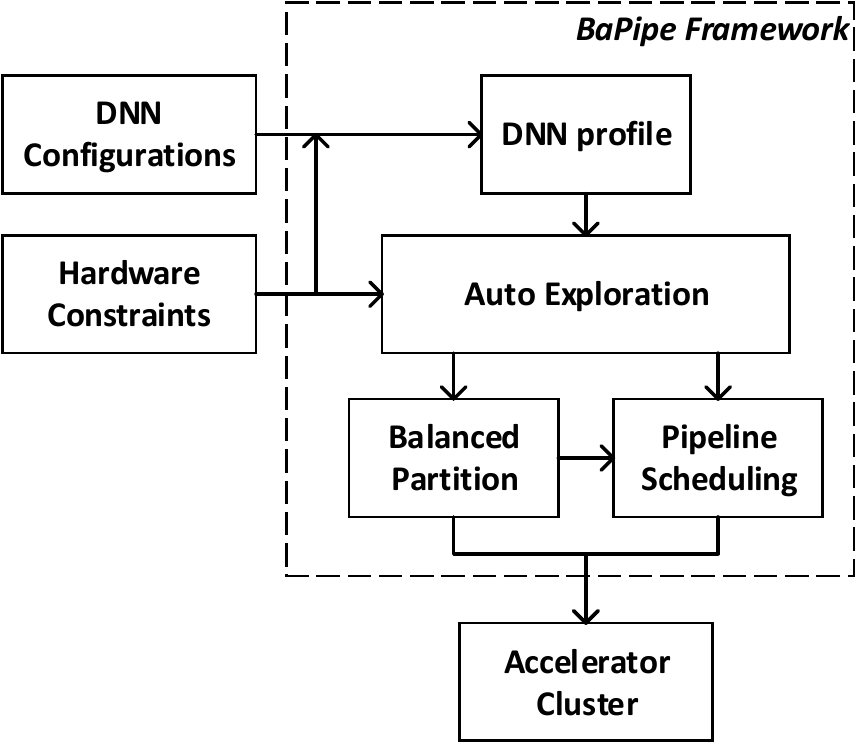}}
\caption{Overview of BaPipe framework.}
\label{framework}
\end{figure}

\section{BaPipe Framework}

\subsection{Overview}

In this section, we present BaPipe, a novel distributed DNN training framework with intra-batch pipeline parallelism. And BaPipe is adaptive to various hardware architectures, including GPU clusters and FPGA clusters in both homogeneous and heterogeneous environment. As shown in Fig. \ref{framework}, the BaPipe framework has two sets of inputs: the DNN configurations and the hardware constraints. The hyper-parameters of DNN to describe network structures form the DNN configurations. The hardware constraints involve computing power, memory bandwidth, memory capacity, and communication bandwidth of each accelerator in the cluster.

There are mainly three parts in the BaPipe framework: DNN profile, automatic exploration of pipeline scheduling, and automatic exploration of balanced partition. BaPipe profiles the DNN first to get computing time for FP and BP, weights size, and features size of every layer in the network. For heterogeneous GPU clusters, BaPipe records the measurement through a short profiling run of 1000 mini-batches on each different type of GPU. For FPGA clusters, BaPipe simulates the DNN profile according to DNN configurations and hardware constraints include computing power and memory bandwidth of each different model of FPGA. And the simulation is based on the hardware accelerator architecture of FPDeep\cite{geng2018fpdeep,geng2018framework,wang2020fpdeep}. Then the results of profiling, the number of accelerators, and memory capacity and communication bandwidth of accelerators are used as the input of both auto exploration methodologies. The partition algorithm partitions the network into several stages across accelerators to balance the computational load, communication cost, and memory consumption. Thus, according to hardware constraints and the results of DNN profiling and partitioning, the pipeline scheduling strategy automatically explores the most suitable sequence of the pipeline. Finally, BaPipe exports the results of pipeline scheduling and balanced partition to deploy the accelerator cluster for parallel DNN training.

\subsection{Pipeline Scheduling}
\begin{figure}[tbp]
%\centerline{\includegraphics[width=\linewidth]{fig/as-s.pdf}}
\centerline{\includegraphics[width=\linewidth]{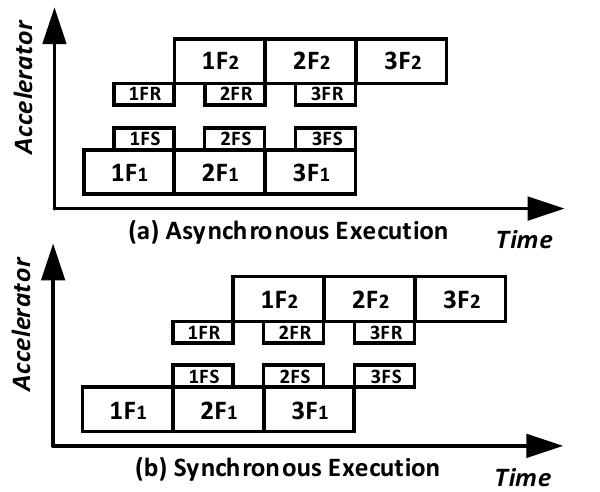}}
\caption{The timeline of a 2-accelerator cluster for DNN inference showing the order of computation and communication: (a) Asynchronous Execution; (b) Synchronous Execution.}
\label{as2s}
\end{figure}

As shown in Fig. \ref{as2s}, there are two execution orders of accelerators for computation and communication: asynchronous execution and synchronous execution. In Fig. \ref{as2s}, mFn means computation time for FP of the m-th batch at accelerator n while mFS/mFR refers to the time to send/receive forward activations of m-th batch between 2 accelerators.

GPUs have to compute and communicate synchronously which means the outputs of computation should be sent after the whole computation. Meanwhile, asynchronous execution can be used in FPGAs, thus communication can start after the part of outputs being done. So the overlap of communication and computation is different between the two execution methods.

\subsubsection{Asynchronous Execution}

\begin{figure*}[tbp]
\centerline{\includegraphics[width=\linewidth]{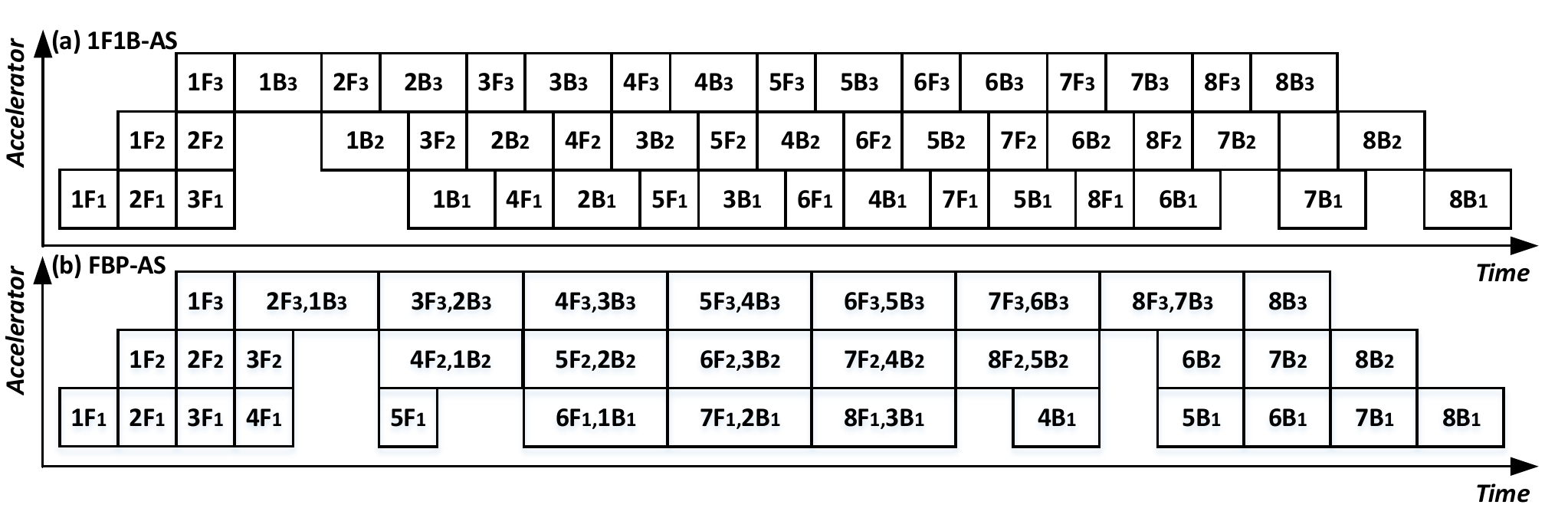}}
\caption{The timeline of a 3-accelerator cluster for DNN training showing the pipeline scheduling methodologies with asynchronous execution: (a) 1F1B-AS;  (b) FBP-AS.}
\label{as}
\end{figure*}

BaPipe provides two options for automatic exploration of pipeline scheduling with asynchronous execution: 1F1B-AS and FBP-AS (Fig. \ref{as}). In Fig. \ref{as} mFn/mBn denotes forward/backward propagation of m-th batch at accelerator n. Moreover, send and receive are omitted due to the complete overlap by asynchronous execution.

\begin{table*}[tbp]
\caption{Comparison between 1F1B-AS and FBP-AS}
\begin{center}
%\resizebox{\linewidth}{!}{
\normalsize
\setlength{\tabcolsep}{7mm}{
\begin{tabular}{c|cc}
 & \textbf{1F1B-AS} & \textbf{FBP-AS} \\
\hline 
\textbf{Mini-batch time} & $(M+N-1)*(F+B)$ & $(M+N-1)*(F+B)$ \\
\textbf{Pipeline bubble} & $\frac{N-1}{M+N-1}$ & $\frac{N-1}{M+N-1}$ \\
\textbf{Features memory} & $(N-i+1)*a$ & $2*(N-i+1)*a$ \\
\textbf{Weights memory} & $2*w$ & $2*w$ \\
\textbf{Bandwidth} & $a/F$ & $2a/(F+B)$ \\
\end{tabular}
}
\label{tab-as}
\end{center}
\end{table*}

1F1B is first proposed by PipeDream\cite{narayanan2019pipedream} which means each accelerator alternately executes forward and backward propagation for its stage in a steady pipeline state. BaPipe modifies it to 1F1B-AS for intra-batch pipeline parallelism with asynchronous scheduling which partitions each mini-batch in several micro-batches, such as 8 micro-batches in Fig. \ref{as}, and updates weights after backward propagation of each mini-batch to ensure the convergence of training and decrease memory requirement of weight stashing. FBP-AS is first used by FPDeep\cite{geng2018fpdeep} which means each accelerator executes forward and backward propagation in parallel with asynchronous scheduling in a steady pipeline state.

The comparison of theoretical performance between 1F1B-AS and FBP-AS is shown in TABLE \ref{tab-as} where $M$ is the number of micro-batches in each mini-batch, $N$ is the number of accelerators in the cluster, $F/B$ are the computation time of FP/BP which are assumed the same in each stage by the balanced partition, $i$ is the i-th accelerator across 1 to N, and $a/w$ is the size of activation/weights in each stage. As we can see, the training time and bubble overhead of each mini-batch, and memory requirements for weights and weights gradient of each stage are the same between 1F1B-AS and FBP-AS when the partition of DNN is the same too. Meanwhile, the activation memory requirement of FBP-AS is twice as much as 1F1B-AS, as the number of micro-batches for FP before the first BP of 1F1B-AS is half of FBP-AS. Besides, the demand bandwidth to overlap communication and computation of 1F1B-AS is more than FBP-AS, since $a$ is the same that send/receive time is the same but the calculation to overlap is different between FP and parallel FP/BP.

It is worth noting that the minimum size of micro-batch to fully utilize DSP resources of FPGA by FP only or parallel FP/BP is different. Thus M in the same size of mini-batch of FBP-AS can be smaller than 1F1B-AS so that the training time is faster, bubble overhead is smaller, and activation memory decreases but bandwidth demand increases. Therefore, BaPipe automatically explores the better asynchronous pipeline scheduling methods for FPGA clusters according to hardware constraints, DNN profile, and DNN partitions.

\begin{figure*}[htbp]
\centerline{\includegraphics[width=\linewidth]{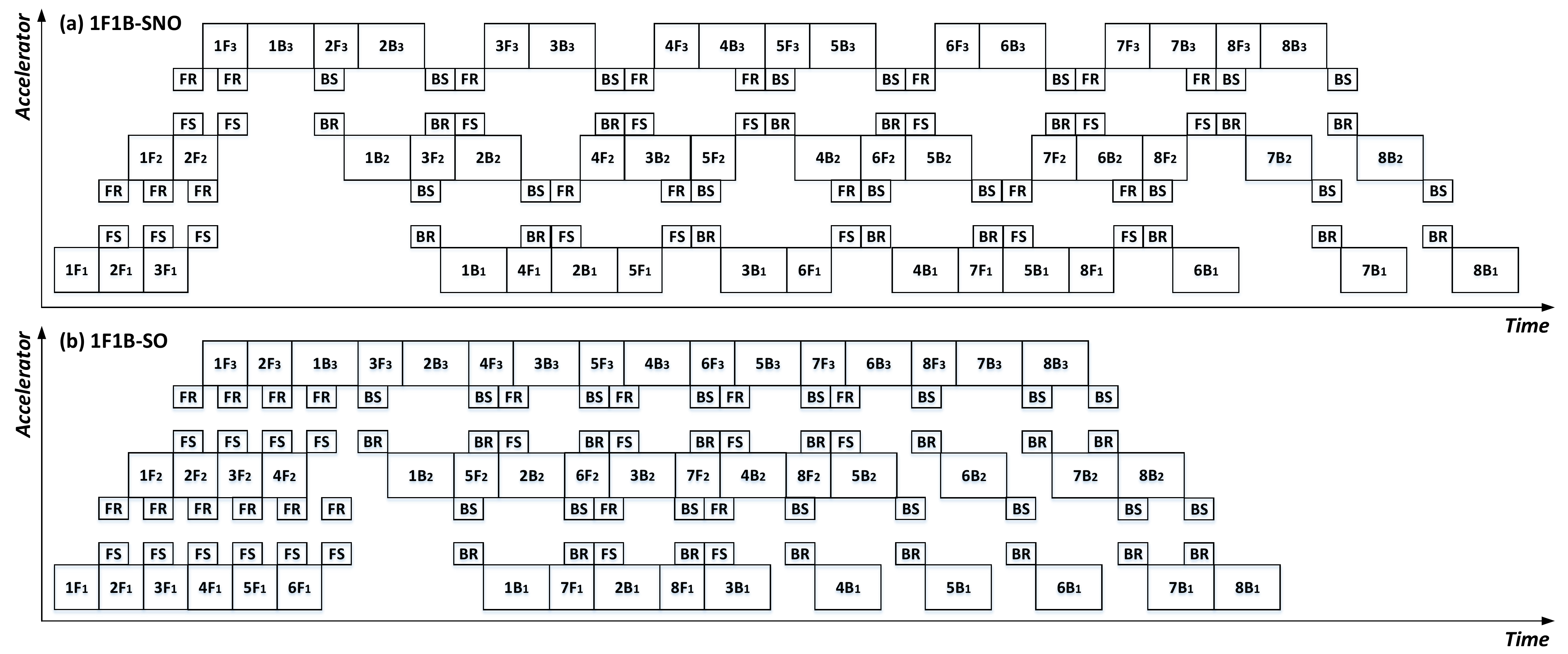}}
\caption{The timeline of a 3-accelerator cluster for DNN training showing the pipeline scheduling methodologies with synchronous execution: (a) 1F1B-SNO;  (b)1F1B-SO.}
\label{syn}
\end{figure*}

\begin{table*}[htbp]
\caption{Comparison between 1F1B-SNO and 1F1B-SO}
\begin{center}
\resizebox{\linewidth}{!}{
\begin{tabular}{c|cc}
 & \textbf{1F1B-SNO} & \textbf{1F1B-SO} \\
\hline 
%\textbf{Mini-batch time} & $(M+N-1)*(F+B)+(N+M-2-\lceil\frac{M-1}{N}\rceil)*2*SR$ & $(M+N-1)*(F+B)+(N-1)*2*SR$ \\
\multirow{2}{*}{\textbf{Mini-batch time}} & $(M+N-1)*(F+B)+$ & $(M+N-1)*(F+B)+$ \\
 & $(N+M-2-\lceil\frac{M-1}{N}\rceil)*2*SR$ & $(N-1)*2*SR$ \\
\textbf{Pipeline bubble} & $\frac{(N-1)*(F+B+2*SR)+(M-1-\lceil\frac{M-1}{N}\rceil)*2*SR}{(M+N-1)*(F+B)+(N+M-2-\lceil\frac{M-1}{N}\rceil)*2*SR}$ & $\frac{(N-1)*(F+B+2*SR)}{(M+N-1)*(F+B)+(N-1)*2*SR}$ \\
\textbf{Features memory} & $(N-i+1)*a$ & $2*(N-i+1)*a$ \\
\textbf{Weights memory} & $2*w$ & $2*w$ \\
\textbf{Bandwidth} & $a/F$ & $a/F$ \\
\end{tabular}
}
\label{tab-syn}
\end{center}
\end{table*}

\subsubsection{Synchronous Execution}

Different tasks such as FP and BP cannot be calculated in parallel by GPU with current machine learning frameworks such as PyTorch\cite{paszke2019pytorch}, Tensorflow\cite{abadi2016tensorflow}, and Caffe\cite{jia2014caffe} due to their sequential instructions. Thus FBP cannot be used by GPUs while 1F1B-AS cannot be used either because of the synchronous execution of GPUs. As shown in Fig. \ref{syn}(a), the naive 1F1B (called 1F1B-SNO in BaPipe) suffers non-overlapping of communication and computation because there are not enough warm-up micro-batches in the set-up pipeline state to completely receive the inputs of each mini-batch before the related computation in each accelerator. In Fig. \ref{syn}, FS/FR denotes the time to send/receive outputs/inputs features in FP while BS/BR denotes the time to send/receive outputs/ inputs error in BP. Also, we assume all input/output activation/error of each stage is the same so that the FS, FR, BS, and BR of each accelerator are all same.

Thus we present 1F1B-SO to overlap the communication and computation of each stage for synchronous pipeline scheduling used in GPU cluster by double the warm-up micro-batches as shown in Fig. \ref{syn}(b). TABLE \ref{tab-syn} shows the comparison of theoretical performance between 1F1B-SNO and 1F1B-SO in where SR denotes the time for send/receive features/errors while other abbreviations are the same as TABLE \ref{tab-as}. As we can see, the training time and bubble overhead of each mini-batch of 1F1B-SO are much better than 1F1B-SNO when the partition of DNN is the same. The extra bubble of 1F1B-SNO consists of non-overlapping communications that are proportional to M. However, the cost of higher performance is double the memory requirements for intermediate features which leads to a decrease of micro-batch size. Moreover, the throughput of training with small batch sizes may be lower when the utilization of GPU is not high enough. Thus the profile of DNN should consider batch size as a variation for GPU clusters by profiling run on each different GPU. After that, BaPipe also automatically explores the better synchronous pipeline scheduling methods for GPU clusters according to hardware constraints, DNN profile, and DNN partitions.

\subsection{Balanced Partition}

To balance the computational load, communication cost, and memory consumption of each accelerator, BaPipe automatically explores the DNN partitioning method by the inter-layer partition with intra-layer partition, coarse-grained partition based on communication bandwidth, and fine-tune partition based on memory capacity, respectively.

The partition algorithm first executes inter-layer partition, assuming that communication and computation are overlap. Second, BaPipe judges whether the communication is the bottleneck, that is, whether the communication time of each stage is longer than the computation time. And if so, BaPipe partitions layers in a coarse-grained manner to overlap communication. Then, BaPipe tunes layer partition if memory capacity is not enough until the bottleneck is resolved. Thus, we can skip the later steps and output the result of DNN partitions. In contrast, if the communication is not the bottleneck, BaPipe skip to the next step. Third, the intra-layer partition is used to make the workload more balanced. Finally, BaPipe judges whether the memory capacity of each accelerator is enough for the requirements of each stage, and if not, finely tunes layer partition until memory requirements are satisfied and communication is not the bottleneck, thus the balanced partition strategy can be output.

\subsubsection{Inter-layer partition}
The input of inter-layer partition for pipeline parallelism is the number of accelerators and computing time for FP and BP of every layer in the network over each different type of accelerator from profiling results. And the inter-layer partition algorithm outputs the DNN partition strategy without considering communication and memory.

Let $T_n$ denote the training time of one micro-batch for the whole network used in accelerator $n$, $l_n$ denotes the part of the network on stage $n$ with the ideal partition so that $L=\sum_{n=1}^N l_n$, where $L$ is the number of layers in the network. If we assume that DNN is uniform across all layers, the ideal training time of each stage is $T=\frac{l_1}{L}*T_1=\frac{l_2}{L}*T_2=\cdots=\frac{l_N}{L}*T_N$, and can be solved as:
\begin{equation}
T=\frac{1}{\sum_{n=1}^N \frac{1}{T_n}}\label{eq}
\end{equation}
Therefore, BaPipe partitions DNN according to T firstly, and then iterates to load balancing with inter-layer partition.

\subsubsection{Intra-layer partition}

The intra-layer partition for pipeline parallelism is first proposed by FPDeep\cite{geng2018fpdeep}. However, FPDeep is used for FPGA clusters in which every FPGA is the same, while we present BaPipe for heterogeneous accelerator clusters (e.g. heterogeneous GPU clusters, heterogeneous FPGA clusters) in which each accelerator can be different. Thus the balanced workload cannot be divided by N directly. The input of intra-layer partition for pipeline parallelism is both the input and output of inter-layer partition. Moreover, the intra-layer partition will be employed only when the communication is not the bottleneck, because intra-layer partition will increase the communication time. BaPipe partitions DNN according inter-layer partition firstly and then iterates to load balancing with intra-layer partition by fine-tuning part of the boundary layer between accelerators.

\subsubsection{Coarse-grained partition based on communication}

When the send/receive time of intermediate activation/error across accelerators is longer than the computation time of each stage, the communication becomes the bottleneck for the pipeline. Since the bandwidth across accelerators is fixed, we can set the feature threshold $a_{th}$ for transmission when the communication time threshold is determined such as T. Thus the DNN can be set to a coarse-grained network in which each layer is consists of consecutive layers in the original DNN between two adjacent layers that the size of their output features both are below $a_{th}$. Therefore, the balanced partition used in the coarse-grained network no longer suffers from a communication bottleneck.

\begin{table*}[tbp]
\caption{Comparison of epoch time among DP, PipeDream, GPipe and Bapipe}
\begin{center}
\resizebox{\linewidth}{!}{
\begin{tabular}{|c|c|c|c|c|c|c|c|c|c|}
\hline
\multirow{2}{*}{Model} & \multirow{2}{*}{Cluster} & \multicolumn{8}{c|}{Epoch time(Speedup over DP)} \\ \cline{3-10} 
 &  & \multicolumn{2}{c|}{DP} & PipeDream\cite{narayanan2019pipedream} & \multicolumn{2}{c|}{GPipe\cite{huang2019gpipe}} & \multicolumn{3}{c|}{Bapipe} \\ \hline
\multirow{4}{*}{VGG-16\cite{simonyan2014very}} & \multirow{2}{*}{4 V100} & B=32 & 0.62x & 1.97x & \multirow{2}{*}{M=8 B=32} & \multirow{2}{*}{1.95x} & \multirow{2}{*}{1F1B-SO} & \multirow{2}{*}{M=32 B=32} & \multirow{2}{*}{2.61x} \\ \cline{3-5}
 &  & B=64 & 1x & 2.12x &  &  &  &  &  \\ \cline{2-10} 
 & \multirow{2}{*}{8 V100} & B=32 & 0.55x & 2.33x & \multirow{2}{*}{M=8 B=32} & \multirow{2}{*}{1.77x} & \multirow{2}{*}{1F1B-SNO} & \multirow{2}{*}{M=64 B=64} & \multirow{2}{*}{2.97x} \\ \cline{3-5}
 &  & B=64 & 1x & 2.94x &  &  &  &  &  \\ \hline
\multirow{2}{*}{ResNet-50\cite{he2016deep}} & 4 V100 & \multicolumn{2}{c|}{1x} & 1x & \multicolumn{2}{c|}{1x} & \multicolumn{3}{c|}{1x} \\ \cline{2-10} 
 & 8 V100 & \multicolumn{2}{c|}{1x} & 1x & \multicolumn{2}{c|}{1x} & \multicolumn{3}{c|}{1x} \\ \hline
\multirow{2}{*}{GNMT-8\cite{wu2016google}} & 4 V100 & B=64 & 1x & 2.75x & M=8 B=64 & 2.93x & 1F1B-SO & M=32 B=64 & 3.90x \\ \cline{2-10} 
 & 8 V100 & B=64 & 1x & 2.37x & M=12 B=64 & 2.05x & 1F1B-SO & M=64 B=64 & 3.21x \\ \hline
\end{tabular}
}
\label{tab-gputime}
\end{center}
\end{table*}

\begin{table*}[tbp]
\caption{Comparison of Maximum (L, W) of GNMT with Other Parallelism}
\begin{center}
%\resizebox{\linewidth}{!}{
\normalsize
\setlength{\tabcolsep}{5mm}{
\begin{tabular}{ccccc}
\hline
GPU Cluster(16GB each) & 1 V100  & 2 V100  & 4 V100   & 8 V100   \\ \hline
DP                     & (32, 445.6M) & (32, 445.6M) & (32, 445.6M)  & (32, 445.6M)  \\
PipeDream              & (32, 445.6M) & (32, 445.6M) & (32, 445.6M)  & (32, 445.6M)  \\
GPipe                  & (32, 445.6M) & (42, 550.6M) & (60, 739.5M)  & (74, 886.4M)  \\
BaPipe                 & (32, 445.6M) & (74, 886.4M) & (118, 1.35B) & (158, 1.78B) \\ \hline
\end{tabular}
}
\label{tab-bigmodel}
\end{center}
\end{table*}

\begin{table}[tbp]
\caption{FPGA Platform Parameters}
\begin{center}
\resizebox{\linewidth}{!}{
\begin{tabular}{ccc}
\hline
Platform & Xilinx VCU118 & Xilinx VCU129 \\ \hline
DSP Slices & 6840 & 12288 \\
On-chip RAM(Mb) & 345.9 & 454.9 \\
GTY/GTM Transceivers & 120/0 & 32/48 \\
DDR4 Throughput(GB/s) & $\sim$40 & $\sim$40 \\ \hline
\end{tabular}
}
\label{tab-FPGA-platform}
\end{center}
\end{table}

\section{Evaluation}
In this section, we evaluate the efficiency of BaPipe with two kinds of DNNs on different accelerator clusters. First, we have trained several small models on GPU clusters to prove performance improvement. And we also evaluate the memory demand with giant models on GPU clusters. Besides, we have simulated the performance of Bapipe on FPGA clusters.

\subsection{Experimental Setup}

We use two representative types of DNN models in our experiments: convolutional model such as VGG-16\cite{simonyan2014very}, ResNet-50\cite{he2016deep} and sequence-to-sequence model such as Google's Neural Machine Translation(GNMT)\cite{wu2016google} with different number of LSTM layers. The GPU cluster we used has 8 NVIDIA V100 GPUs, with 16GB of GPU device memory, and PCIe Gen3 x16 interconnections between GPUs. The FPGA clusters we used in the simulator have 4 Xilinx VCU118 boards and 4 Xilinx VCU129 boards as shown in TABLE \ref{tab-FPGA-platform}.

\subsection{Experimental Results on GPU clusters}

\subsubsection{Small DNN models experiments}\label{sec-small}

TABLE \ref{tab-gputime} compares several DNN models training epoch time among DP, PipeDream\cite{narayanan2019pipedream}, GPipe\cite{huang2019gpipe} and our work. We use the GLOO\cite{gloo} communication backend for all parallel training as the NCCL does not currently support multi-threads communication in safety, and we set DP as a baseline for all parallel training. In TABLE \ref{tab-gputime}, B means the amount of batch size per GPU, and to improve performance we set B as much as possible under the constraint of GPU memory. As we can see, the throughput of DNN training has reduced when B has decreased in DP and PipeDream. For GPipe and BaPipe, M means the number of micro-batches in each mini-batch, and we also set M large enough to ignore the pipeline bubble. But the number of micro-batches is still limited by the GPU memory for GPipe, as the accelerator in GPipe has to store all activation of a whole mini-batch. So GPipe selects to recompute forward propagation for memory reduction, and in our experiments, we didn’t use recomputation for fairness. Also, we use the same partitions as BaPipe in GPipe, since GPipe doesn’t have a load balancing algorithm, and PipeDream uses its strategy to partition training load.

As TABLE \ref{tab-gputime} shown, BaPipe has up to 3.2x faster than DP, 1.4x faster than state-of-the-art pipeline parallelism in DNN training. And except for VGG-16 on 8 V100, the pipeline scheduling of BaPipe automatically chose 1F1B-SNO, to overlap communication and computation for more performance improvement than the utilization reduction caused by mini-batch decreased in other situations. For ResNet-50, both BaPipe and PipeDream have explored that the best partition is DP on our GPU cluster, as the communication consumption for activation between GPUs in the pipeline is larger than the amount of communication for weights gradient among GPUs in DP.

\subsubsection{Giant DNN models experiments}

TABLE \ref{tab-bigmodel} compares the maximum model size which DP, PipeDream, GPipe, and BaPipe can support on different GPU clusters. We use the GNMT-L model to evaluate memory consumption. GNMT-L is based on GNMT\cite{wu2016google} which consists of an LSTM model. Thus, we have stacked the GNMT-L by L/2 encoder and L/2 decoder layers, and W presents the number of parameters. We set the batch size per GPU is 32, and the number of micro-batches in each mini-batch for GPipe and BaPipe both are two times the number of pipeline stages. As mentioned in Section \ref{sec-small}, the GPipe in our evaluation does not support recomputation. And BaPipe has used 1F1B-SNO to fit the biggest model size for DNN training on clusters.

As shown in TABLE \ref{tab-bigmodel}, the model size is constrained by single GPU memory limits with DP and PipeDream because of weight stashing for convergence. On the other hand, GPipe and BaPipe both can scale DNN model size with the number of accelerators increased. BaPipe can train up to 4x larger models than DP and 2x larger models than state-of-the-art pipeline parallelism.

\begin{table}[tbp]
\caption{Comparison between BaPipe and DP on FPGA clusters}
\begin{center}
\resizebox{\linewidth}{!}{
\begin{tabular}{cccc}
\hline
ResNet-50                                                              & 4 VCU118 & 2 VCU129, 2 VCU118 & 4 VCU129 \\ \hline
\begin{tabular}[c]{@{}c@{}}Batch time\\ (Speedup over DP)\end{tabular} & 1x       & 1.05x             & 1.14x    \\ \hline
\end{tabular}
}
\label{tab-resnet-FPGA}
\end{center}
\end{table}

\subsection{Experimental Results on FPGA clusters}

TABLE \ref{tab-resnet-FPGA} compares the simulation result of ResNet-50 training on different FPGA clusters between DP and BaPipe, and the parameters of FPGA boards are shown as TABLE \ref{tab-FPGA-platform}. We set 1 as the micro-batch size for BaPipe and 128 as the mini-batch size for both BaPipe and DP. We use fp16 precision for memory optimizer, and guarantee weights of each stage are stored in on-chip memory as much as possible for BaPipe. At the same time, DP has to store weights in DDR due to the size limits, so there has speedup over DP compared with TABLE \ref{tab-gputime}. However, the computation resources of FPGAs are not enough to take advantage of on-chip memory, thus, the speedup is only 1.14x. Also, BaPipe automatically chooses FBP-AS to increase accelerator utilization for clusters in the simulator.

\section{Conclusion}
This paper presents BaPipe, a pipeline parallelism training framework, which explores the design space of pipeline scheduling methods and balanced partition strategies for DNN training on accelerator clusters. Compared to state-of-the-art parallelism frameworks, BaPipe provides up to 3.2x speedup and 4x memory reduction in various platforms. Consequently, BaPipe can train small DNN models more efficiently and train far larger DNN models.

\section*{Acknowledgment}
This research was sponsored by Huawei Innovation Research Program. The numerical calculations in this paper have been done on the supercomputing system in the Supercomputing Center of University of Science and Technology of China. Supports from Chio-In IEONG are gratefully acknowledged.
%\printbibliography
\bibliographystyle{IEEEtran}
\bibliography{ref}

\end{document}